\documentclass{elsart}
\usepackage{graphicx}
\usepackage{amssymb}

\begin{document}

\begin{frontmatter}
\title{Stabilization of three--dimensional light bullets by a transverse
lattice in a Kerr medium with dispersion management}
\author[UW]{Micha\l{}~Matuszewski},
\author[IPJ]{Eryk~Infeld},
\author[TAU]{Boris~A.~Malomed},
\author[UW]{Marek~Trippenbach}
\address[UW]{Institute of Theoretical Physics, Physics Department,
Warsaw University, Ho\.{z}a 69, PL-00-681 Warsaw, Poland}
\address[IPJ]{Soltan Institute for Nuclear Studies,
Ho\.{z}a 69, PL-00-681 Warsaw, Poland}
\address[TAU]{Department of Interdisciplinary Sciences,
School of Electrical Engineering,
Faculty of Engineering, Tel Aviv University, Tel Aviv 69978, Israel}

\begin{abstract}
We demonstrate a possibility to stabilize three--dimensional
spatiotemporal solitons (``light bullets") in self--focusing Kerr
media by means of a combination of dispersion management in the
longitudinal direction (with the group--velocity dispersion
alternating between positive and negative values) and periodic
modulation of the refractive index in one transverse direction,
out of the two. The analysis is based on the variational
approximation (results of direct three-dimensional simulations
will be reported in a follow-up work). A predicted stability area
is identified in the model's parameter space. It features a
minimum of the necessary strength of the transverse modulation of
the refractive index, and finite minimum and maximum values of the
soliton's energy. The former feature is also explained
analytically.
\end{abstract}

\end{frontmatter}

\section{Introduction}

Search for spatiotemporal solitons in diverse optical media, alias
``light bullets" (LBs) \cite{Yaron}, is a challenge to fundamental
and applied research in nonlinear optics, see original works
\cite{KR,Koshiba,chi2,Miriam,tandem,Frank,Isaac,saturable,Wagner}
and a very recent review \cite{review}. Stationary solutions for
LBs can easily be found in the cubic ($\chi ^{(3)}$)
multi-dimensional nonlinear Schr\"{o}dinger (NLS) equation
\cite{Yaron}, but their stability is a problem, as they are
unstable against spatiotemporal collapse \cite{Berge}. The problem
may be avoided by resorting to milder nonlinearities, such as
saturable \cite{saturable}, cubic-quintic \cite{CQ}, or quadratic
($\chi ^{(2)}$) \cite{KR,Koshiba,chi2,Miriam,tandem,Frank,Isaac}.

Despite considerable progress in theoretical studies, three-dimensional (3D)
LBs in a bulk medium have not yet been observed in an experiment. The only
successful experimental finding reported thus far was a stable quasi-2D
spatiotemporal soliton in $\chi ^{(2)}$ crystals \cite{Frank} (the
tilted-wavefront technique \cite{Paolo}, used in that work, precluded
achieving self-confinement in one transverse direction). On the other hand,
it was predicted \cite{Isaac} that a spatial cylindrical soliton may be
stabilized in a bulk medium composed of layers with alternating signs of the
Kerr coefficient. Similar stabilization was then predicted for what may be
regarded as 2D\ solitons in Bose-Einstein condensates (BECs), with the
coefficient in front of the cubic nonlinear term subjected to periodic
modulation in time via the \textit{Feshbach resonance} in external ac
magnetic field \cite{FR,Castilla}. However, no stable 3D soliton could be
found in either realization (optical or BEC) of this setting.

Serious difficulties encountered in the experimental search for
LBs in 3D media is an incentive to look for alternative settings
admitting stable\emph{\/} 3D optical solitons. With the Kerr
nonlinearity, a possibility is to use a layered structure that
periodically reverses the sign of the local group-velocity
dispersion (GVD), without affecting the $\chi ^{(3)}$ coefficient.
This resembles a well-known scheme in fiber optics, known as
\textit{dispersion management}\ (DM), see, e.g., Refs. \cite{DM}
and review \cite{Progress}. A 2D generalization of the DM scheme
was recently proposed, assuming a layered planar waveguide of this
type, uniform in the transverse direction \cite{we,Spain}. As a
result, large stability regions for the 2D spatiotemporal solitons
were identified, including double-peaked breathers; however, a 3D
version of the same model could not give rise to any stable
soliton \cite{we}. It was also shown in Ref. \cite{we} that no
stable 3D soliton could be found in a more sophisticated model,
which combines the DM and periodic modulation of the Kerr
coefficient in the longitudinal direction.

Another approach to the stabilization of multidimensional solitons
was developed in the context of the self-attracting BEC. It is
based on the corresponding Gross-Pitaevskii equation which
includes a periodic potential created as an optical lattice (OL,
i.e., an interference pattern produced by illuminating the
condensate by counter-propagating coherent laser beams). It has
been demonstrated that 2D \cite{BBB,Yang,Estoril} and 3D
\cite{BBB} solitons can be easily stabilized by the OL of the same
dimension. Moreover, stable solitons can also be readily supported
by \emph{low-dimensional\/} OLs, i.e., 1D and 2D ones in the 2D
\cite{Estoril,BBB2} and 3D \cite{Estoril,BBB2,Barcelona} cases,
respectively; additionally, a 3D soliton can be stabilized by a
cylindrical (\textit{Bessel}) lattice \cite{Dumitru}, similar to
one introduced, in the context of 2D models, in Ref.
\cite{Bessel}. On the other hand, 3D solitons cannot be stabilized
by a 1D periodic potential \cite{BBB2}; however, the 1D lattice
potential in combination with the above-mentioned time-periodic
modulation of the nonlinearity, provided by the Feshbach resonance
in the ac magnetic field, supports single- and multi-peaked stable
3D solitons in vast areas of the respective parameter space
\cite{we-new}.

The above results suggest a possibility of existence of stable 3D
``bullets" in a $\chi ^{(3)}$ medium with the DM in the
longitudinal direction ($z$), additionally equipped with an
effective lattice potential (i.e., periodic modulation of the
refractive index) in one transverse direction ($y$), while in the
remaining transverse direction ($x$) the medium remains uniform.
If this is possible, stable LBs will be definitely possible too in
a medium with the periodic modulation of the refractive index in
both transverse directions; however, the setting with one uniform
direction is more interesting in terms of steering solitons and
studying collisions between them \cite{Estoril,BBB2}. The
objective of the present work is to predict such 3D spatiotemporal
solitons and investigate their stability. Our first consideration
of this possibility is based on the variational approximation
(VA); systematic simulations of the 3D model are quite
complicated, and will be presented in a follow-up work. It is
relevant to mention that the existence and stability of 3D
solitons in the Gross-Pitaevskii equation with the quasi-2D
periodic potential, which were originally predicted by the
VA\cite{Estoril,BBB2}, was definitely confirmed by direct
simulations \cite{Estoril,BBB2,Barcelona}, which suggests that in
the present model the 3D solitons may easily be stable too.

The model is based on the normalized NLS equation describing the evolution
of the local amplitude $u$ of the electromagnetic wave, which is a
straightforward extension of the 2D model put forward in Ref. \cite{we}:
\begin{equation}
i\frac{\partial u}{\partial z}+\left[ \frac{1}{2}\left(
\frac{\partial ^{2}}{\partial x^{2}}+\frac{\partial ^{2}}{\partial
y^{2}}+D(z)\frac{\partial ^{2}}{\partial \tau ^{2}}\right)
+\varepsilon \cos (2y)+|u|^{2}\right] u=0. \label{general}
\end{equation}Here, $\varepsilon $ is the strength of the transverse modulation (the
modulation period is normalized to be $\pi $), while $\tau $ and $D(z)$ are
the same reduced temporal variable and local GVD coefficient as in the
fiber-optic DM models \cite{DM,Progress}.

Equation (\ref{general}) implies the propagation of a linearly polarized
wave, with the single component $u$; a more general situation will be
described by a two-component (vectorial) version of Eq. (\ref{general}),
with the two polarization coupled, as usual, by the cubic
cross-phase-modulation terms. We do not expect that the vectorial model will
produce results qualitatively different form those presented below. As
usual, the NLS equation assumes the applicability of the paraxial
approximation, i.e., the spatial size of solitons (see below) must be much
larger than the underlying wavelength of light, which is definitely a
physically relevant assumption \cite{review}, and the temporal part of the
equation implies that the higher-order GVD is negligible (previous
considerations have demonstrated that the higher-order dispersion does not
drastically alter DM solitons \cite{TOD}).

As is commonly adopted, we assume a symmetric \textit{DM map}, with equal
lengths $L$ of the normal- and anomalous-GVD segments (usually, the results
are not sensitive to the map's asymmetry),
\begin{equation}
D(z)=\left\{
\begin{array}{l}
\overline{D}+D_{\mathrm{m}}>0,\,0<z<L, \\
\overline{D}-D_{\mathrm{m}}<0,\,L<z<2L,\end{array}\right.
\label{D(z)}
\end{equation}the average dispersion being much smaller than the
modulation amplitude,
$\left\vert \overline{D}\right\vert \ll D_{\mathrm{m}}$. Using the scaling
invariances of Eq. (\ref{general}), we fix $L\equiv 1$ and
$D_{\mathrm{m}}\equiv 1$.

Recently, a somewhat similar 2D model was introduced in Ref.
\cite{Salerno}. The most important difference is that it has the
variable coefficient $D(z)$ multiplying \emph{both} the GVD and
diffraction terms, $u_{\tau \tau }$ and $u_{xx}$. Actually, that
model was motivated by a continuum limit of some discrete systems;
in the present context, it would be quite difficult to implement
the periodic reversal of the sign of the transverse diffraction.

\section{The variational approximation}

Aiming to apply the VA for the search of LB solutions (a review of the
variational method can be found in Ref. \cite{Progress}), we adopt the
Gaussian \textit{ansatz},
\begin{eqnarray}
u &=&A(z)\exp \left\{ \mathrm{i}\phi (z)-\frac{1}{2}\left[
\frac{x^{2}}{W^{2}(z)}+\frac{y^{2}}{V^{2}(z)}+\frac{\tau
^{2}}{T^{2}(z)}\right] \right. +
\nonumber \\
&&+\left. \frac{\mathrm{i}}{2}\left[ b(z)\,x^{2}+c(z)\,y^{2}+\beta (z)\,\tau
^{2}\right] \right\} ,  \label{ansatz}
\end{eqnarray}where $A$ and $\phi $ are the amplitude and phase of the soliton,
$T$ and $W,V$ are its temporal and two transverse spatial widths, and $\beta $
and $b,c$ are the temporal and two spatial chirps. The Lagrangian from which
Eq. (\ref{general}) can be derived is
\begin{eqnarray}
L &=&\frac{1}{2}\int_{-\infty }^{+\infty }\,\mathrm{d}x\int_{-\infty
}^{+\infty }\,\mathrm{d}y\int_{-\infty }^{+\infty }\,\mathrm{d}\tau \left[
\mathrm{i}\left( u_{z}u^{\ast }-u_{z}^{\ast }u\right) -\left\vert
u_{x}\right\vert ^{2}-\left\vert u_{y}\right\vert ^{2}-D\left\vert u_{\tau
}\right\vert ^{2}\right. \\
&&\left. +2\varepsilon \cos (2y)|u|^{2}+|u|^{4}\right] .
\end{eqnarray}The substitution of the ansatz (\ref{ansatz}) in this expression and
integrations lead to an \textit{effective Lagrangian}, with the prime
standing for $d/dz$:
\begin{eqnarray}
(4/\pi ^{3/2})L_{\mathrm{eff}} &=&A^{2}WVT\left[ 4\phi ^{\prime }-b^{\prime
}W^{2}-c^{\prime }V^{2}-\beta ^{\prime }T^{2}-W^{-2}-V^{-2}-DT^{-2}\right.
\nonumber \\
&&\left. -b^{2}W^{2}-c^{2}V^{2}+\varepsilon \exp \left( -V^{2}\right)
-D(z)\beta ^{2}T^{2}+A^{2}/\sqrt{2}\right] ,  \label{effL}
\end{eqnarray}

The first variational equation, $\delta L_{\mathrm{eff}}/\delta \phi =0$,
applied to Eq. (\ref{effL}) yields the energy conservation, $dE/dz=0$, with
\begin{equation}
E\equiv A^{2}WVT.  \label{E}
\end{equation}The conservation of $E$ is used to eliminate $A^{2}$ from the set of
subsequent equations, $\delta L_{\mathrm{eff}}/\delta \left( W,V,T,b,c,\beta
\right) =0$. They can be arranged so as, first, to eliminate the chirps,
\begin{equation}
b=W^{\prime }/W,\,c=V^{\prime }/V,\beta =D^{-1}T^{\prime }/T.  \label{betab}
\end{equation}the remaining equations for the spatial and temporal widths being

\begin{eqnarray}
W^{\prime \prime } &=&\frac{1}{W^{3}}-\frac{E}{2\sqrt{2}W^{2}VT},
\label{variat1} \\
V^{\prime \prime } &=&\frac{1}{V^{3}}-4\varepsilon V\exp \left(
-V^{2}\right) -\frac{E}{2\sqrt{2}WV^{2}T},  \label{variat2} \\
\left( \frac{T^{\prime }}{D}\right) ^{\prime }
&=&\frac{D}{T^{3}}-\frac{E}{2\sqrt{2}WVT^{2}}.  \label{variat3}
\end{eqnarray}{The Hamiltonian of these equations, which is a dynamical invariant in the
case of constant $D$, is
\[
{\mathcal{H}}=\left( W^{\prime }\right) ^{2}+\left( V^{\prime
}\right) ^{2}+\frac{\left( T^{\prime }\right)
^{2}}{D}+\frac{1}{W^{2}}+\frac{1}{V^{2}}+\frac{D}{T^{2}}-4\varepsilon
\exp (-V^{2})-\frac{E}{\sqrt{2}WVT}
\]} In the case of the piece-wise constant modulation, such as in (\ref{D(z)}),
the variables $W$, $W^{\prime }$, $V$, $V^{\prime }$, $T$ and $\beta $
must be continuous at junctions between the segments with $D_{\pm }\equiv
\overline{D}\pm D_{\mathrm{m}}$. As it follows from Eq. (\ref{betab}), the
continuity of the temporal chirp $\beta (z)$ implies a jump of the
derivative $T^{\prime }$ when passing from $D_{-}$ to $D_{+}$, or vice
versa:
\begin{equation}
\left( T^{\prime }\right) _{D=D_{+}}=\left( D_{+}/D_{-}\right) \left(
T^{\prime }\right) _{D=D_{-}}.  \label{jump}
\end{equation}

In the case of a continuous DM map, rather than the one
(\ref{D(z)}), Eq. (\ref{variat3}) has a formal singularity at the
points where $D(z)$ vanishes, changing its sign. However, it is
known that there is no real singularity in this case, as
$T^{\prime }$ vanishes at the same points, which cancels the
singularity out \cite{Progress}.

In the absence of the DM and transverse modulation, i.e., $D\equiv
+1$ and $\varepsilon =0$, three equations (\ref{variat1}) -
(\ref{variat3})\ reduce to one, which is tantamount to the
variational equation derived in Ref. \cite{Sweden} from the
spatiotemporally isotropic ansatz [cf. Eq. (\ref{ansatz})],
$u=A\exp \left[ \mathrm{i}\phi -(1/2)\left(
W^{-2}+\mathrm{i}b\right) \left( x^{2}+y^{2}+\tau ^{2}\right)
\right] $. In particular, this single equation correctly predicts
the asymptotic law of the \textit{strong collapse} in the 3D case,
which is stable against anisotropic perturbations \cite{Russia},
$V=W=T\approx \left( 5E/3\sqrt{2}\right) ^{1/5}\left(
z_{0}-z\right) ^{2/5}$, $z=z_{0}$ being the collapse point. The
location of this point is determined by initial conditions, but,
in any case, it belongs to an interval $D>0$, where the GVD is
anomalous.

Another possible collapse scenario is an effectively
two-dimensional (weak) one, with two widths shrinking to zero as
$z_{0}-z\rightarrow 0$, while the third one remains finite. For
instance, the corresponding asymptotic law may be\begin{equation}
V=T=A\left( z_{0}-z\right)
^{1/2},~W=\frac{\sqrt{2}E}{4+A^{4}}-\frac{\left( 4+A^{4}\right)
^{2}}{4\sqrt{2}EA^{2}}\left( z_{0}-z\right) \ln \left(
z_{0}-z\right) ,  \label{2Dcollapse}
\end{equation}where $A$ is a positive constant or else $V=2/T\sim (z_{0}-z)^{1/2}$
and $W\rightarrow W_{0}$ (in this case too, the collapse point
$z_0$ must belong to a segment with $D>0$). In direct simulations
of Eqs. (\ref{variat1}) - (\ref{variat3}), we actually observed
only the latter scenario. However, we did not specially try to
find initial conditions that could initiate a solution
corresponding to the strong 3D collapse, as our objective is not
the study of the collapse, but rather search for solitons stable
against collapse. In fact, known results for the solitons in the
3D Gross-Pitaevskii equation with the OL potential suggest that,
while the VA may be incorrect in the description of the collapse,
as a singular solution, it provides for quite accurate predictions
for the stability of solitons as \emph{regular solutions}
\cite{Estoril,BBB2}.

If the DM is absent, and the constant GVD is normal, i.e.,
$D\equiv -1$, only the 2D collapse in the transverse plane would
be possible, so that (cf. Eq. (\ref{2Dcollapse}))\[ V=W=A\left(
z_{0}-z\right) ^{1/2},~T=\frac{\sqrt{2}E}{4+A^{4}}+\frac{\left(
4+A^{4}\right) ^{2}}{4\sqrt{2}EA^{2}}\left( z_{0}-z\right) \ln
\left( z_{0}-z\right) .
\]However, we did not observed this collapse scenario in our simulations. The
same comment as one given above pertains to this case as well.

A possibility of the stabilization of the 3D soliton by a
sufficiently strong lattice can be understood noticing that, for
large $\varepsilon $, one may keep only the first two terms on the
right-hand side of Eq. (\ref{variat2}). This approximation yields
a nearly constant value $V_{0}$ of $V$, which is a smaller root of
the corresponding equation,
\begin{equation}
4\varepsilon V_{0}^{4}\exp \left( -V_{0}^{2}\right) =1  \label{static}
\end{equation}(a larger root corresponds to an unstable equilibrium).
The two roots exist provided that
\begin{equation}
\varepsilon >\varepsilon _{\mathrm{\min }}=\mathrm{e}^{2}/16\approx
\allowbreak 0.46,  \label{min}
\end{equation}the relevant one being limited by $V_{0}<2$.
Then, the substitution of $V=V_{0}$ in the remaining
equations (\ref{variat1}) and (\ref{variat3})
leads to essentially the same VA-generated dynamical system as
derived for the 2D DM model in Ref. \cite{we}, which was shown to
give rise to stable spatiotemporal solitons. On the other hand, it
was demonstrated in Ref. \cite{we} too that, in the case of
$\varepsilon =0$, the 3D VA equations, as well as the full
underlying 3D model, have no stable soliton solutions.

The stabilization of the LB in the present model for large
$\varepsilon $ can also be understood in a different way, without
resorting to VA: in a very strong lattice, the soliton is trapped
entirely in a single ``valley" of the periodic potential, and the
problem thus reduces to a nearly 2D one, where spatiotemporal
solitons may be stable, cf. a similar stabilization mechanism for
the solitons in the Gross-Pitaevskii equations developed in
\cite{Castilla}. From this point of view, a really interesting
issue is to find an \emph{actual} minimum $\varepsilon _{\min }$
of the lattice's strength which is necessary for the stabilization
of the 3D solitons, as at $\varepsilon $ close enough to
$\varepsilon _{\mathrm{\min }} $ the stabilized solitons are truly
3D objects, rather than their nearly-2D counterparts.

\section{Results}

\begin{figure}[tbp]
\includegraphics[width=13.5cm]{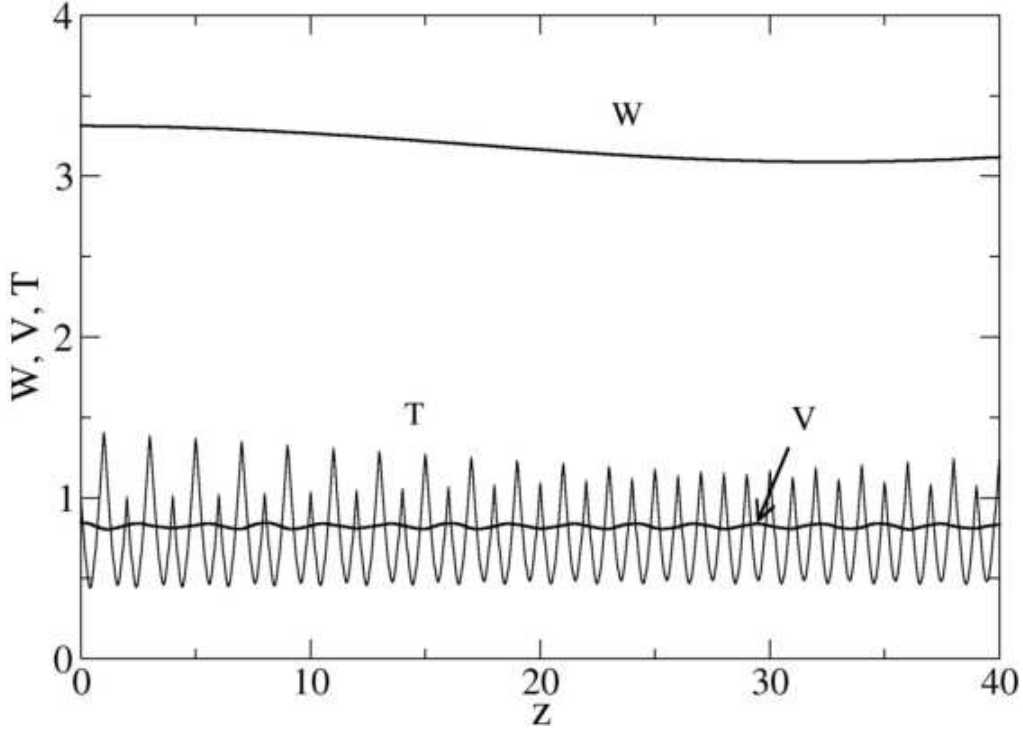}
\caption{An example of the stable evolution of solutions to the
variational equations (\protect\ref{variat1}) -
(\protect\ref{variat3}). The soliton's widths in the direction
$x,y$ and $\protect\tau $, i.e., $W,V$ and $T$, are shown as
functions of $z$, for $E=0.5$, $\protect\varepsilon =1$, and
$\overline{D}=0$.} \label{fig1}
\end{figure}

We explored the parameter space of the variational system
(\ref{variat1}) - (\ref{variat3}), $\left( E,\varepsilon
,\overline{D}\right) $, by means of direct simulations of the
equations (with regard to the jump condition (\ref{jump})). As a
result, it was possible to identify regions where the model admits
\emph{stable} solitons featuring regular oscillations in $z$ with
the DM-map period. An example of such a regime is shown in Fig.
\ref{fig1} (oscillations in the evolution of $W$ are not visible
in the figure because, as an estimate demonstrates, their
amplitude is $\simeq 0.001$).

\begin{figure}[tbp]
\includegraphics[width=13.5cm]{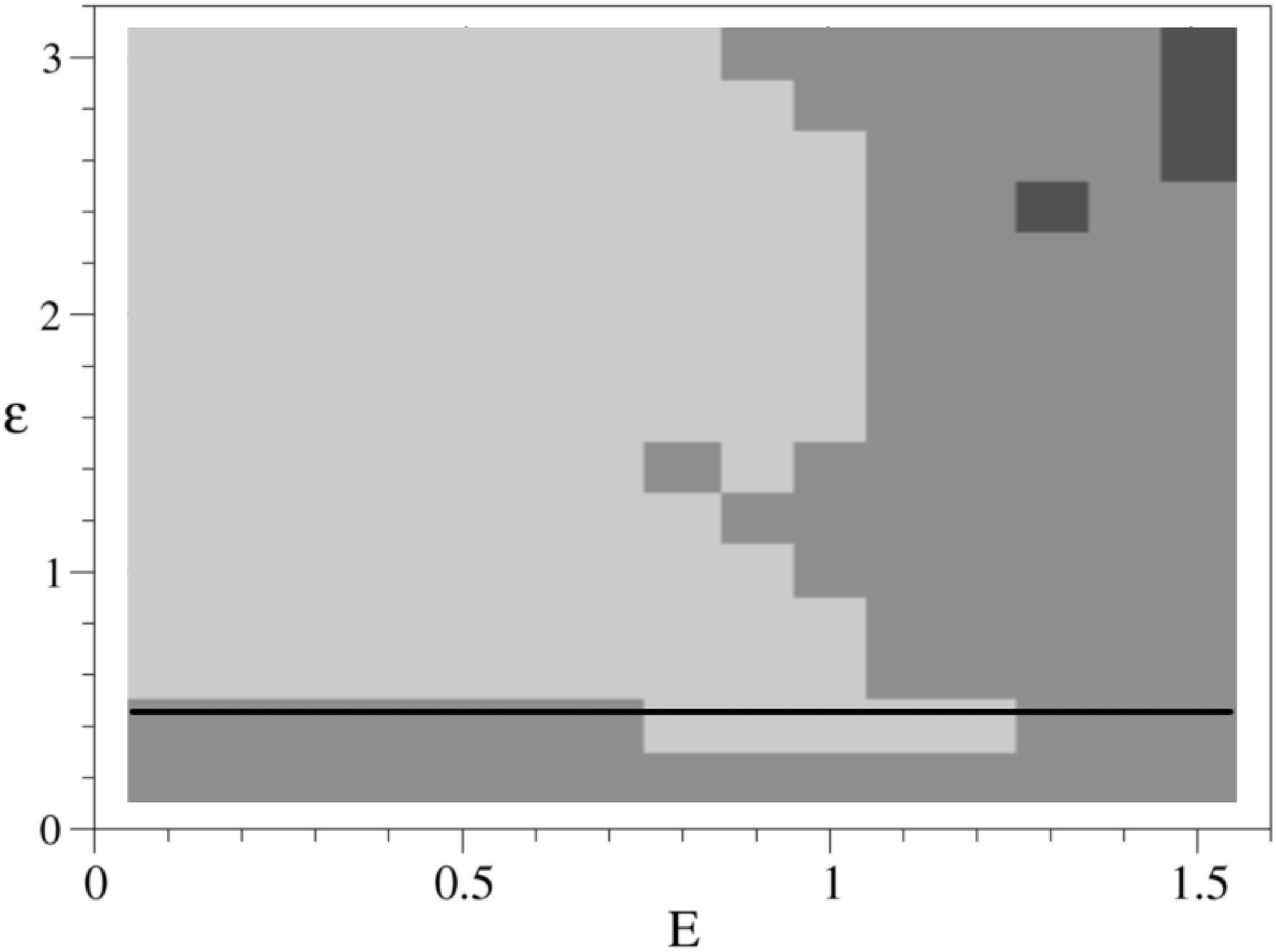}
\caption{The stability area for the 3D spatiotemporal solitons in
the $\left( E,\protect\varepsilon \right) $ plane, with
$\overline{D}=0$, is shown by light-gray shading. In gray and
dark-gray regions, the 3D soliton is predicted, respectively, to
spread out and collapse. The vertical line corresponds to the
analytically predicted threshold (\protect\ref{min}).}
\label{fig2}
\end{figure}

\begin{figure}[tbp]
\includegraphics[width=13.5cm]{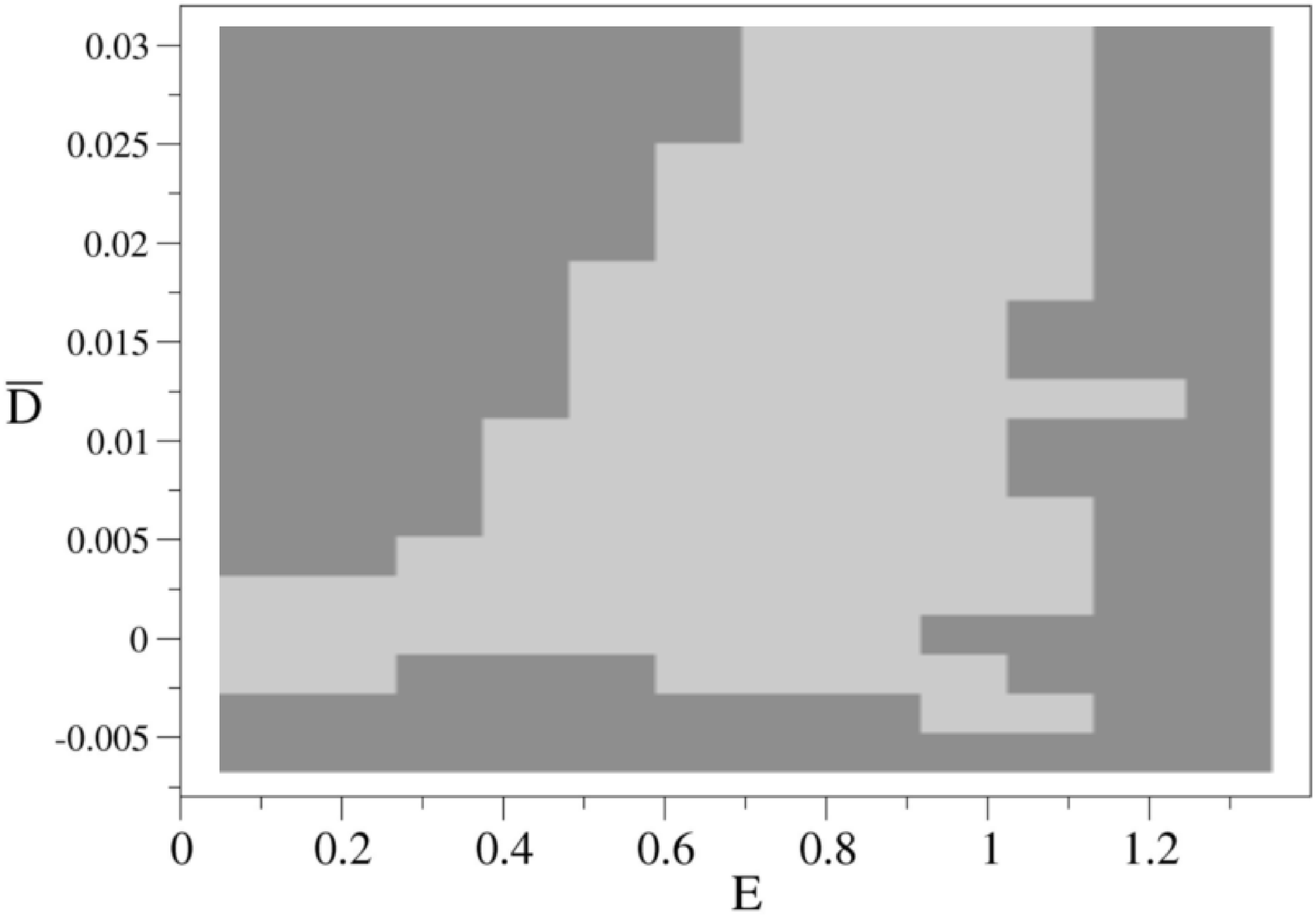}
\caption{The stability area in the $\left( E,\overline{D}\right) $ plane,
with $\protect\varepsilon =1$, is shown by light--gray shading. In the gray
region, the 3D soliton is predicted to spread out.}
\label{fig3}
\end{figure}

Systematic results obtained from the simulations are summarized in stability
diagrams displayed in Figs. \ref{fig2} and \ref{fig3}. A remarkable fact,
apparent in Fig. \ref{fig2}, is that the minimum value of the lattice's
strength, $\varepsilon _{\min }=0.46$, at which the solitons may be stable,
coincides with the analytical prediction (\ref{min}), up to the available
numerical accuracy.

The existence of a maximum value $E_{\max }$ of the energy
admitting the stable LBs is, essentially, a quasi-2D feature,
which can be understood assuming that the potential lattice is
strong. Indeed, as explained above, in such a case the value of
$V$ is approximately fixed as the smaller root of Eq.
(\ref{static}). Within a segment where the GVD coefficient keeps
the constant value, $D_{+}=\overline{D}+D_{\mathrm{m}}>0$, which
corresponds to anomalous dispersion (see Eq. (\ref{D(z)}), the
remaining equations (\ref{variat1}) and (\ref{variat3}) are
tantamount to those for a uniform 2D Kerr-self-focusing medium,
hence the energy is limited by the value $E_{\max }$ corresponding
to the \textit{Townes soliton}; the soliton will collapse if
$E>E_{\max }$ \cite{Berge}.

The fact that the region of stable solitons is also limited by a
minimum energy, $E_{\min }$, except for the case of
$\overline{D}=0$, when $E_{\min }=0$ (see Fig. \ref{fig3})), is
actually a quasi-1D feature, which is characteristic to the DM
solitons in optical fibers. In that case, the term $\sim E$ in the
evolution equation for $T(z)$, cf. Eq. (\ref{variat3}), is
necessary to balance the average GVD coefficient $\overline{D}$,
so that $E_{\min }$ and $\overline{D}$ vanish simultaneously
\cite{DM}. It is noteworthy too that, as well as in the case of
the 1D DM solitons in fibers, the stability area in Fig.
\ref{fig3} includes a part with \emph{normal} average GVD,
$\overline{D}<0$, which seems counterintuitive, but can be
explained \cite{DM}. This part extends in Fig. \ref{fig3} up to
$\left( -\overline{D}\right) _{\max }\approx 0.005$.


\section{Conclusions}

In this work, we have proposed a possibility to stabilize
spatiotemporal solitons (``light bullets") in three-dimensional
self-focusing Kerr media by means of the dispersion management
(DM), which means that the local group-velocity dispersion
coefficient alternates between positive and negative values along
the propagation direction, $z$. Recently, it was shown that the DM
alone can stabilize solitons in 2D (planar) waveguides, but in the
bulk (3D) DM medium the ``bullets" are unstable. In this work, we
have demonstrated that the complete stabilization can \ be
provided if the longitudinal DM is combined with periodic
modulation of the refractive index in one transverse direction
($y$), out of the two. The analysis was based on the variational
approximation (systematic results of direct simulations will be
reported in a follow-up paper). A stability area for the light
bullets was identified in the model's parameter space. Its salient
features are a necessary minimum strength of the transverse
modulation of the refractive index, and minimum and maximum values
$E_{\min ,\max }$ of the soliton's energy. The former feature can
be accurately predicted (see Eq. (\ref{min})) in an analytical
form from the evolution equation for the width of the soliton in
the $y$-direction. The existence of $E_{\min }$, which vanishes
when we assume zero average dispersion, can be explained in the
same way as for the temporal solitons in DM optical fibers. Also,
similar to the case of DM solitons in fibers, we find that the
stability area extends to a region of \emph{normal} average
dispersion \cite{DM}. On the other hand, the existence of $E_{\max
}$ can be understood similarly to as it was recently done in the
2D counterpart of the present model (the strong transverse lattice
can squeeze the system to a nearly 2D shape).

The results presented in this work suggest a new approach to the challenging
problem of the creation of 3D spatiotemporal optical solitons. The model
also opens a way to address advanced issues, such as collisions between the
LBs, and the existence and stability of solitons with different symmetries
(for instance, LBs which are odd in the longitudinal and/or transverse
directions). These issues will be considered elsewhere.

\section{Acknowlegdements}

M.M., M.T. and E.I. acknowledge support from KBN Grant No. 2P03 B4325.
B.A.M. acknowledges the hospitality of the Physics Department and Soltan
Institute for Nuclear Studies at the Warsaw University, and partial support
from the Israel Science Foundation grant No. 8006/03. This author also
appreciates the help of A. Desyatnikov in making Ref. \cite{Estoril}
available on the internet.

\end{document}